\documentclass{PoS}
\usepackage{amsmath}
\newcommand{\url}[1]{\href{#1}{#1}}
\title{Porting DDalphaAMG solver to K computer}

\ShortTitle{Porting DDalphaAMG solver to K computer}

\author{%
Ken-Ichi Ishikawa and
\speaker{Issaku Kanamori}\\
        Department of Physical Science, Hiroshima University\\
        Higashi-Hiroshima, Hiroshima 739-8526, Japan\\
        E-mail: \email{ishikawa@theo.phys.sci.horoshima-u.ac.jp},
        \email{kanamori@hiroshima-u.ac.jp}}

\abstract{
We port Domain-Decomposed-alpha-AMG solver to the K computer.
The system has 8 cores and 16 GB memory per node, of which theoretical
peak is 128 GFlops (82,944 nodes in total).
Its feature, as many as 256 registers per core
and as large as 0.5 byte/Flop ratio, requires a different tuning
from other machines.
In order to use more registers, we change some of the data structure
and rewrite matrix-vector operations with intrinsics.
The performance is improved by more than a factor two for twelve solves
including the setup.
The efficiency is still about 5\% after the optimization, which is lower
than a previously tuned mixed precision solver for the K computer, 22\%.
The throughput is, however,
more than two times better for a physical point configuration.
}

\FullConference{The 36th Annual International Symposium on Lattice Field Theory - LATTICE2018\\
		22-28 July, 2018\\
		Michigan State University, East Lansing, Michigan, USA.}

\begin{document}

\section{Introduction}

The performance bottle neck of lattice QCD simulations is linear solvers
for the Dirac equations in most cases.
Liner solvers are used both in generating configuration with HMC
algorithm and in measuring physical observables which contain fermion propagators.
As the system becomes closer to the continuum limit and the fermion mass
becomes lighter, the Dirac operator
becomes singular
because the smallest eigenvalue in magnitude
approaches zero (in lattice unit).
The computational cost for the iterative solvers diverges, known as
a critical slowing down.
Depending on the algorithm of the iterative solver, one even may not obtain
the convergent result with the state of arts configurations of which
pion mass is (nearly) physical point.
Therefore
it is very important to accelerate the solvers to make
the simulations as close as the real world to suppress
the systematic uncertainties:
the light quark mass is as light as the
physical quark, the lattice volume is large and the lattice spacing is small.

One of great progress in accelerating the solver in the last decade
is the application of multigrid algorithm \cite{Brannick:2007ue, Babich:2010qb}.
The multigrid solver can drastically reduce the above mentioned critical
slowing down.
Since QCD is a gauge theory, naive geometric multigrid methods do not work.
A combination of aggregation based algebraic multigrid
and an adaptive setup, which is often called adaptive multigrid,
 is the right way.
It is in fact a hybrid of the geometric and algebraic method.
The domain decomposed aggregation is a geometrical blocking procedure,
and the procedure to build a coarse grid operator by using (low mode)
vectors is exactly the algebraic method.
The key feature of the algorithm is the adaptive setup to make the vectors
we have just mentioned.
It allows us to build a coarse grid operator which
captures the low mode space of the original fine grid operator very
efficiently.
Therefore the solution from the coarse grid solver provides a good
approximation to the low mode part of the whole system.
The high mode part is solved on the fine grid (smoother).
A combination of these two solvers can be used as 
a preconditionor of any iterative outer solvers.
One of the merits of the multigrid method is that it is fast even with
small quark mass. 
Another merit is that the coarse grid solver can recursively have a coarse
grid which can further accelerate the solver.
A demerit is that the adaptive setup requires a
significant computational cost.

The adaptive multigrid algorithm for QCD was first applied
to Wilson \cite{Babich:2010qb} and Clover fermions \cite{Osborn:2010mb}.
It is then applied to Domainwall fermions \cite{Cohen:2012sh}
and recently to staggered fermions \cite{Brower:2018ymy}. 
Depending on the choice of solvers at each step, there are variations
of the algorithm.
In this work, we focus on the DD$\alpha$AMG algorithm
for clover fermions \cite{Frommer:2013fsa}.
It uses a domain decomposed solver for the smoother process
and is rather close to
the domain decomposed inexact deflation algorithm \cite{Luscher:2007se}.
The DD$\alpha$AMG is also used as a preconditionor for overlap fermion
\cite{Brannick:2014vda},
and is extended to twisted mass fermion \cite{Alexandrou:2016izb}.
An implementation of adaptive multigrid solvers on GPU with QUDA is
also available \cite{7877146}.
Another implementation for many core SIMD machines with GRID
is presented in this conference \cite{Daniel}.

 \section{Algorithm}
\label{sec:algorithm}

We follow the implementation \texttt{DDalphaAMG} available
at \url{https://github.com/DDalphaAMG}, which we refer the original code
in the following.
The algorithm is an adaptive aggregation based domain decomposed
algebraic multigrid method \cite{Frommer:2013fsa}.
In this work, we restrict ourselves on the 2-level method.

\begin{figure}

 \centering
 \begin{picture}(300,130)(0,0)
  \put(10,12){\includegraphics[width=280pt]{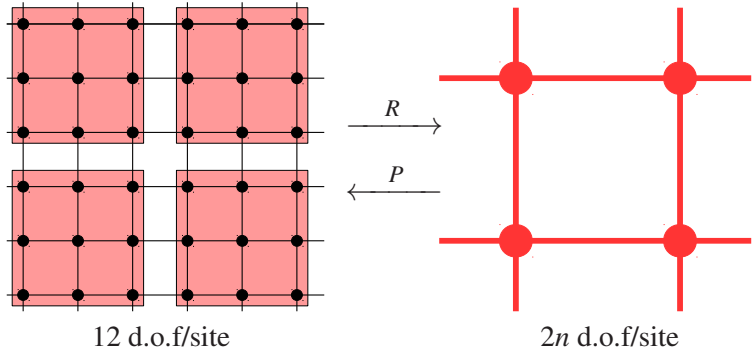}} 
  \put(42,0){\makebox{12 d.o.f/site}}
  \put(210,0){\makebox{$2n$ d.o.f/site}}
  \put(137,80){\makebox{\large$\xrightarrow{\quad R\quad }$}}
  \put(137,55){\makebox{\large$\xleftarrow{\quad P\quad }$}}
 \end{picture}%

 \caption{Mapping from a fine grid lattice (left) to a coarse grid
 lattice (right).
 The fine lattice sites in a domain is mapped to a single coarse
 lattice site.  Here, the degrees of freedom (d.o.f) is for a fermion field
 and $n$ is the number of the test vectors.
 }
 \label{fig:fine_to_coarse}
\end{figure}

Here is a sketch of the algorithm.
The multigrid process is used as a preconditionor for the
outer iterative solver.
Before running the solver, we first prepare
test vectors (also called null-space vectors) $|\lambda_i\rangle$ ($i=1,...,n;\
n\sim 20$), which are rich in low mode of the
Dirac operator.
They are generated with an (iterative) adaptive method.
The first set of $|\lambda_i\rangle$ is generated by applying
an approximate solver on random vectors.
Once the initial set is ready,
we can iteratively repeat the following two steps, 1.\ build a multigrid preconditionor $M$ (see below)
with the current $|\lambda_i\rangle$,
2.\ apply (a part of) the multigrid solver
to $|\lambda_i\rangle$ and obtain updated $|\lambda_i\rangle$ with the
low mode contamination enhance.
The updated set of $|\lambda_i\rangle$ is used in the next iteration.
The number of the iterations is a tunable parameter and we use 4 times
iteration throughout in this work.
Having prepared the test vectors, 
we project them onto a given domain $X$ and chirality $s$,
\begin{equation}
 |{\lambda_i (X,s=\pm)}\rangle
  \equiv
  \begin{cases}
     \frac{1\pm \gamma_5}{2} |\lambda_i\rangle & \text{in domain $X$},\\
   0 & \text{otherwise}.
  \end{cases}
\end{equation}
They are used to build a coarse grid operator $D_{\mathrm{coarse}}$ from the fine grid
Dirac operator $D$ as
\begin{equation}
 D \rightarrow
 D_{\mathrm{coarse}}(X,s,i;Y,t,j)
  =\langle{\lambda_i(X,s)}|D|\lambda_j(Y,t)\rangle.
\end{equation}
Here the domain $X$ plays a role of lattice site on the coarse grid,
the lattice sites in the domain are mapped to a single site on the
coarse grid (Fig.~\ref{fig:fine_to_coarse}).
The chirality $s$ is the ``spin'' and the label $i$ is the ``color''
degrees of freedom on the coarse grid so that we have a $2n$ component
vector on each coarse site $X$ for the quark field.
The effective gauge link which connects neighboring sites
on the coarse grid is a $2n\times 2n$ matrix,
which in fact has both ``spin'' and ``color'' degrees of freedom.
A $2n \times 2n$ matrix on the coarse site $X$,
$D_{\mathrm{coarse}}(X,s,i;X,t,j)$, is  ``clover'' term.
The projection of the source vector $|{b}\rangle$ (restriction $R$)
and prolongation of
the solution vector $x_{\mathrm{coarse}}$ on the coarse grid ($P$)
are defined as
\begin{align}
R\colon & |b\rangle
  \rightarrow b_{\mathrm{coarse}}(X,s,i) = \langle \lambda_i(X,s)|b\rangle, \\
P\colon &  x_{\mathrm{coarse}}(X,s,i)
 \rightarrow
 |x\rangle
 = \sum_{X,s,i} x_{\mathrm{coarse}}(X,s,i)|\lambda_i(X,s)\rangle.
\end{align}
To reduce the errors in the high mode, we use another solver
called smoother, $D_{\mathrm{smoother}}^{-1}$.
We use a post smoother only.
The preconditionor matrix $M$ is in the end
\begin{equation}
 M= P D_{\mathrm{coarse}}^{-1} R
   + D_{\mathrm{smoother}}^{-1} ( 1- D P D_{\mathrm{coarse}}^{-1} R).
\end{equation}

We need three solvers in total:
the outer solver, the smoother $D_{\mathrm{smoother}}^{-1}$,
and the coarse grid solver $D_{\mathrm{coarse}}^{-1}$.
The outer solver of the \texttt{DDalphaAMG} is Flexible GMRES.
The smoother is a multiplicative Schwartz Alternating Procedure (SAP),
for which the inner solver in the domain is an site even-odd
preconditioned Minimal Residual solver with fixed number of iteration.
The coarse grid solver is an even-odd preconditioned GMRES solver, of
which the convergence condition for the residual vector is $5\times 10^{-2}$.
As emphasized in \cite{Frommer:2013fsa},
the SAP efficiently reduces the errors of the high
mode of the Dirac operator
while the coarse grid solver reduces the errors of the low mode.
In total, errors from both high and low modes are reduced efficiently
in the multigrid steps.

 \section{K computer and Details of the Tuning}
\label{sec:K-and-tuning}
 
The target system for performance tuning is the K computer
at the Center for Computational Science,
 RIKEN (R-CCS).
The system is available since 2012 but is still one of the leading
HPC machines in Japan.
It is the 3rd in the latest (June, 2018) HPCG ranking and was the first
until one before (November, 2017).
The CPU is the SPARK64 VIIIfx processor equipped with 8 cores, of which
theoretical peak performance is 128 GFlops.
Each node consists of one processor and 16 GB memory, and the
inter-node network is the Tofu interconnection.
The system has 82,944 nodes in total and
the total theoretical peak performance is about 10 PFlops.
 The key features of the K computer for performance tuning are:
 \begin{itemize}
  \item as many as 256 registers/core, which are 128-bit wide
  \item double precision SIMD arithmetics are available (no single
	precision)
  \item relatively large (0.5 byte/FLOP) byte-per-flop ratio
 \end{itemize}
 Since the processor does not have single precision SIMD arithmetics,
 a 128-bit wide SIMD register can not treat 4 single precision numbers
 but only 2 double precision numbers, i.e.,
 1 complex number in double precision.
 Still, however, using single precision numbers can save the memory
 bandwidth and can accelerate the computation.

 For the fine grid operator,
 we rewrite the hopping terms and multiplication of the clover
 term with intrinsics.
 The data structure for the (inverse) clover term is also modified to make use of
 its hermitian property.
 The original code stores the inverse of the clover term as an LU
 decomposition, but our code stores the whole inverse as a hermitian
 matrix as follow.
  In the chiral representation,
 we have 2 hermitian matrices of which size is $6\times 6$.
 We modify the code to store the matrix with a minimal data size:
 6 real
 and 15 complex numbers instead of 36 complex numbers for a $6\times 6$
 hermitian matrix.
 Since we have enough number of registers, all of the 6 real and 15
 complex numbers together with input and output vectors, both are
 made of 6 complex numbers, are stored on the registers simultaneously.
 These changes of the treatment of the (inverse) clover term has the
 biggest impact to the performance in our tuning.

 The coarse grid operator multiplication consists of mainly
 matrix-vector multiplications --- $2n\times 2n$ matrix
 for the effective gauge links and the ``clover'' terms.
 We use $4\times 4$ tiling in these matrix-vector multiplications,
 which are implemented with intrinsics.
 The 4 component vector in the tiling
 has 2-component ``spin'' and 2-component ``color'' degrees of freedom.
 Because of this implementation, the number of the test vectors,
 which is the total number of the ``color'', must be even. 
  We could use a larger tiling size which
 would accelerate the performance more, but would reduce flexibility
 on the choice of the size of the coarse grid operator, or would require
 a more complicated implementation to treat the data which does not fit
 in the tile.

Our performance tuning is implemented only up to 2-level multigrid method.
The original code also has an SSE optimized version but we started with
the unoptimized version.
The code is available at 
  \url{https://github.com/i-kanamori/DDalphaAMG/tree/K/}.

\section{Benchmark Results}
\label{sec:results}

\begin{figure}
\noindent\hfil
 \includegraphics[width=0.46\linewidth]{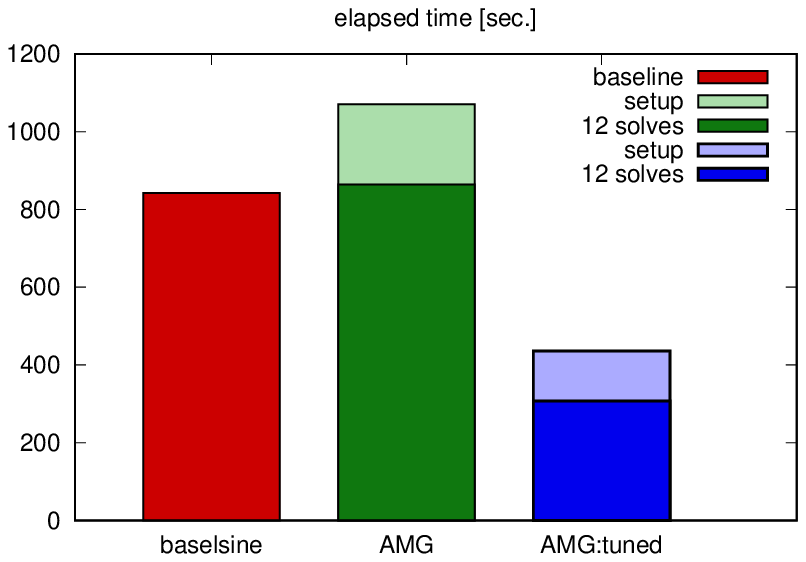} 
\hfil
 \includegraphics[width=0.46\linewidth]{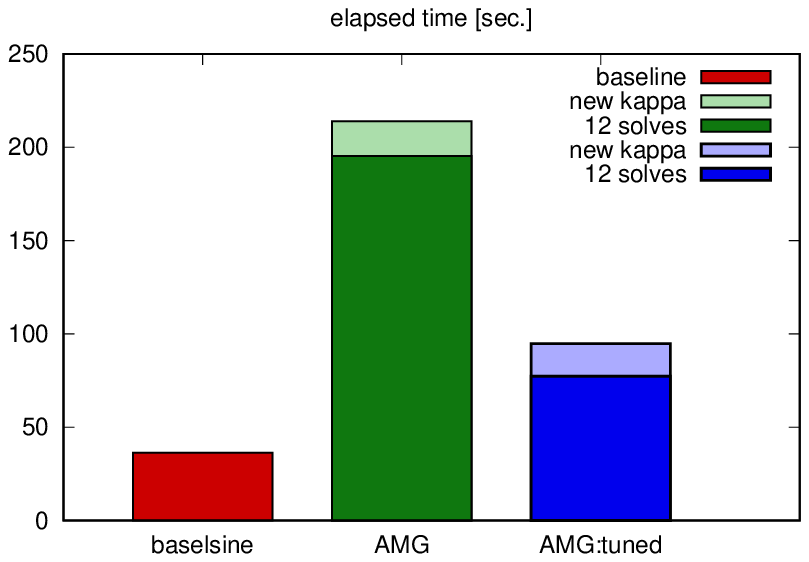}

 \caption{Timing of the solvers for light quark (left panel) and strange
 quark (right panel).
 The baseline (red) is a well-tuned existing solver for the
 K computer \cite{Ishikawa:2015rho}, AMG (green) is the
 \texttt{DDalphaAMG}
 and AMG:tuned (blue) is this work after a performance tuning of
 \texttt{DDalphaAMG}. 
 }
\label{fig:timing} 
\end{figure}

We use a single PACS configuration at almost physical pion mass
$m_\pi \simeq 146\,{\mathrm{MeV}}$ on a $96^4$ lattice generated on
the K computer~\cite{Ishikawa:2015rho}.
The strange quark mass is set to the physical value.
We use 1024 nodes and the local lattice volume is
$12\times 12 \times 12 \times 24$.  The timings for 12 solves
together with
the setup time for
a light quark and a strange quark are plotted in Fig.~\ref{fig:timing}.

In the figure, we also put the result with the same solver as used in
\cite{Ishikawa:2015rho} for comparison (denoted as ``baseline'').
It is a mixed precision nested BiCGstab, where the single precision solver uses
domain decomposition with block size $12\times 12 \times 12 \times 12$
and $N_{\mathrm{SAP}}=5$.
The solver inside the domain uses SSOR method with sub-blocking.
Eight threads coming from 8 cores parallelize inside the domain.
This is a well-tuned solver on the K computer: its efficiency is about
22\%\footnote{
The performance in this proceedings are measured by the hardware counter,
of which run contains contractions for meson spectroscopy and data I/O.
The theoretical value of the efficiency is about
10--20\% lower than the values cited here.}.

For both the original and the tuned version of \texttt{DDalphaAMG}, 
we use $4\times 4\times 4\times 4$ block size
and $N_{\mathrm{SAP}}=4$ for the SAP iterations.  The number of the test
vectors is 16.
With this setting,
the local lattice size on the fine grid
$12\times 12 \times 12 \times 24$ becomes $3\times 3\times 3\times 6$
on the coarse grid, and the degrees of freedom on the site change
from $12$ to $32$.
Inside the domain is not thread parallelized
but
the loops over domains are parallelized.

Let us focus on the timing for the light quark first.
The 12 solves without the setup stage by the original code
 (green one in the figure) spend almost the same time
as the baseline (red one).  The efficiency is, however, much
lower: only 3.0\%.
That is,
the performance with inefficient code 
with the multigrid algorithm is competing
with the well tuned code.
Including the setup (pale green part), 
the multigrid solver is slower at this stage.
After the performance tuning, which is plotted with blue color in the
figure,  the elapsed time becomes about the half of the baseline.
The efficiency of the tuned code is 5.3\%.
To obtain a further improvement of the efficiency,
we would need a drastic change of the code.
For the strange quark, the baseline code is fast enough.
The overhead for making coarse grid operator etc.\ is not
compensated,
even after the tuning.
Here, we use the same test vectors as the light quark so that we can save
the cost for the setup but the coarse grid
operator must be reconstructed.
The ``new kappa'' in the plot (pale colors) is the timing needed
to reconstruct the coarse grid operator
 with a hopping parameter for the strange quark, $\kappa_s$.

\begin{figure}
 \noindent\hfil
 \includegraphics[width=0.46\linewidth]{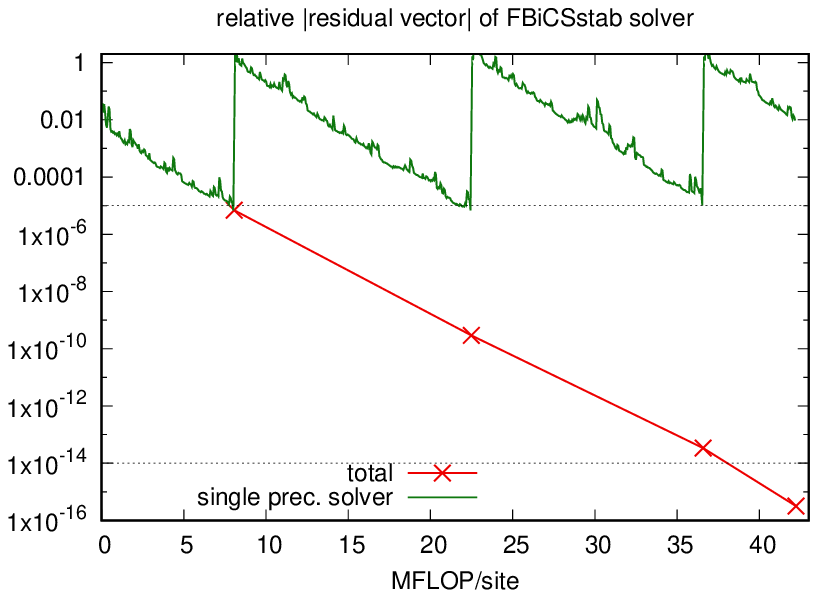} 
\hfil
 \includegraphics[width=0.46\linewidth]{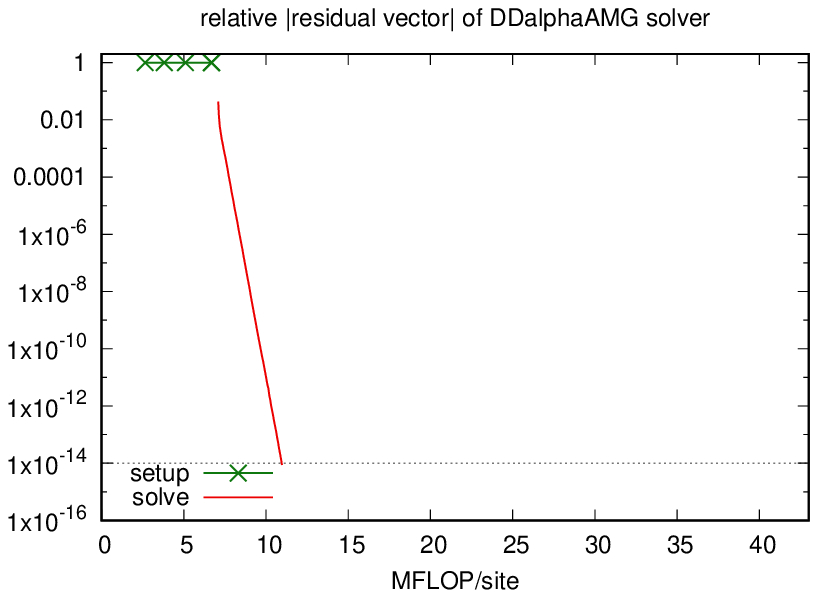}

 \caption{The relative residual norm against the theoretical FLOP
 counting for a light quark,
 with the baseline solver from \cite{Ishikawa:2015rho} (left panel)
 and tuned \texttt{DDalphaAMG} (right panel).
 For the former, the residual norm by both inner single precision solver
 (green solid) and the outer solver (red dashed) are plotted.
 The target tolerance is $10^{-14}$.
 }
\label{fig:flop}

\end{figure}

Figure~\ref{fig:flop} shows the residual norm against
the theoretical counting of FLOPs for the light quark.
We set the same scale for the both panels to see how fast
the multigrid solver is.  For the baseline (left panel),
the residual norms from both the inner single precision solver
and the outer solver are plotted.
The counting is 42 MFLOP/site (504 FLOP/site for 12 solves).
For the tuned version of the \texttt{DDalphaAMG} (right panel),
FLOPs for the setup stage are plotted as well.
It takes 6.7 MFLOP/site for the setup and 4.3 MFLOP/site for a solve
(58.3 MFLOP/site for 1 setup and 12 solves).
For the solving itself, the tuned version of \texttt{DDalphaAMG}
is about 10 times faster.
This is consistent with the timing and efficiency, 
\texttt{DDalphaAMG} is 2.5 times faster in the timing but 4 times slower in
the efficiency.

\section{Summary}
\label{sec:summary}

We ported the implementation of DD$\alpha$AMG solver
(\texttt{DDalphaAMG})
for clover fermions to the K computer and improved the performance.
The efficiency is still 4 times lower than a well-tuned existing solver for the
target machine but the throughput is 2 times better for 12 solves
including the setup stage for almost physical light quarks.
The theoretical FLOP counting showed the multigrid method requires
10 times less FLOP in the solving stage.
For the strange quark, the multigrid solver is slower than the well-tuned
one without multigrid method.

\subsection*{Acknowledgements}

We thank the developers of the original code, especially
we thank M.~Rottmann for helping to run in on the K computer.
This work is supported by Priority Issue 9 to be Tackled by Using Post K
Computer.
We also thanks the Joint Institute for Computational Fundamental
Science(JICFuS) and
the computational resource on the K computer (hp170229, hp180178).

\bibliography{lattice2018proc}

\providecommand{\href}[2]{#2}\begingroup\raggedright\begin{thebibliography}{10}

\bibitem{Brannick:2007ue}
J.~Brannick, R.~C. Brower, M.~A. Clark, J.~C. Osborn and C.~Rebbi,
  \emph{{Adaptive Multigrid Algorithm for Lattice QCD}},
  \href{https://doi.org/10.1103/PhysRevLett.100.041601}{\emph{Phys. Rev. Lett.}
  {\bfseries 100} (2008) 041601}
  [\href{https://arxiv.org/abs/0707.4018}{{\ttfamily 0707.4018}}].

\bibitem{Babich:2010qb}
R.~Babich, J.~Brannick, R.~C. Brower, M.~A. Clark, T.~A. Manteuffel, S.~F.
  McCormick et~al., \emph{{Adaptive multigrid algorithm for the lattice
  Wilson-Dirac operator}},
  \href{https://doi.org/10.1103/PhysRevLett.105.201602}{\emph{Phys. Rev. Lett.}
  {\bfseries 105} (2010) 201602}
  [\href{https://arxiv.org/abs/1005.3043}{{\ttfamily 1005.3043}}].

\bibitem{Osborn:2010mb}
J.~C. Osborn, R.~Babich, J.~Brannick, R.~C. Brower, M.~A. Clark, S.~D. Cohen
  et~al., \emph{{Multigrid solver for clover fermions}},
  \href{https://doi.org/10.22323/1.105.0037}{\emph{PoS} {\bfseries LATTICE2010}
  (2010) 037} [\href{https://arxiv.org/abs/1011.2775}{{\ttfamily 1011.2775}}].

\bibitem{Cohen:2012sh}
S.~D. Cohen, R.~C. Brower, M.~A. Clark and J.~C. Osborn, \emph{{Multigrid
  Algorithms for Domain-Wall Fermions}},
  \href{https://doi.org/10.22323/1.139.0030}{\emph{PoS} {\bfseries LATTICE2011}
  (2011) 030} [\href{https://arxiv.org/abs/1205.2933}{{\ttfamily 1205.2933}}].

\bibitem{Brower:2018ymy}
R.~C. Brower, M.~A. Clark, A.~Strelchenko and E.~Weinberg, \emph{{Multigrid
  algorithm for staggered lattice fermions}},
  \href{https://doi.org/10.1103/PhysRevD.97.114513}{\emph{Phys. Rev.}
  {\bfseries D97} (2018) 114513}
  [\href{https://arxiv.org/abs/1801.07823}{{\ttfamily 1801.07823}}].

\bibitem{Frommer:2013fsa}
A.~Frommer, K.~Kahl, S.~Krieg, B.~Leder and M.~Rottmann, \emph{{Adaptive
  Aggregation Based Domain Decomposition Multigrid for the Lattice Wilson Dirac
  Operator}}, \href{https://doi.org/10.1137/130919507}{\emph{SIAM J. Sci.
  Comput.} {\bfseries 36} (2014) A1581}
  [\href{https://arxiv.org/abs/1303.1377}{{\ttfamily 1303.1377}}].

\bibitem{Luscher:2007se}
M.~Luscher, \emph{{Local coherence and deflation of the low quark modes in
  lattice QCD}},
  \href{https://doi.org/10.1088/1126-6708/2007/07/081}{\emph{JHEP} {\bfseries
  07} (2007) 081} [\href{https://arxiv.org/abs/0706.2298}{{\ttfamily
  0706.2298}}].

\bibitem{Brannick:2014vda}
J.~Brannick, A.~Frommer, K.~Kahl, B.~Leder, M.~Rottmann and A.~Strebel,
  \emph{{Multigrid Preconditioning for the Overlap Operator in Lattice QCD}},
  \href{https://doi.org/10.1007/s00211-015-0725-6}{\emph{Numer. Math.}
  {\bfseries 132} (2016) 463}
  [\href{https://arxiv.org/abs/1410.7170}{{\ttfamily 1410.7170}}].

\bibitem{Alexandrou:2016izb}
C.~Alexandrou, S.~Bacchio, J.~Finkenrath, A.~Frommer, K.~Kahl and M.~Rottmann,
  \emph{{Adaptive Aggregation-based Domain Decomposition Multigrid for Twisted
  Mass Fermions}},
  \href{https://doi.org/10.1103/PhysRevD.94.114509}{\emph{Phys. Rev.}
  {\bfseries D94} (2016) 114509}
  [\href{https://arxiv.org/abs/1610.02370}{{\ttfamily 1610.02370}}].

\bibitem{7877146}
M.~A. Clark, B.~Jo{\'o}, A.~Strelchenko, M.~Cheng, A.~Gambhir and R.~C. Brower,
  \emph{{Accelerating Lattice QCD Multigrid on GPUs Using Fine-Grained
  Parallelization}},  in \emph{SC '16: Proceedings of the International
  Conference for High Performance Computing, Networking, Storage and Analysis},
  pp.~795--806, Nov, 2016, \href{https://arxiv.org/abs/1612.07873}{{\ttfamily
  1612.07873}}, \href{https://doi.org/10.1109/SC.2016.67}{DOI}.

\bibitem{Daniel}
D.~Richtmann, T.~Wettig and P.~Peter~Boyle, \emph{{Multigrid for Wilson Clover
  Fermions in Grid}}, {\emph{talk in this conference} (2018) }.

\bibitem{Ishikawa:2015rho}
{\scshape PACS} collaboration, K.~I. Ishikawa, N.~Ishizuka, Y.~Kuramashi,
  Y.~Nakamura, Y.~Namekawa, Y.~Taniguchi et~al., \emph{{2+1 Flavor QCD
  Simulation on a $96^4$ Lattice}},
  \href{https://doi.org/10.22323/1.251.0075}{\emph{PoS} {\bfseries LATTICE2015}
  (2016) 075} [\href{https://arxiv.org/abs/1511.09222}{{\ttfamily
  1511.09222}}].

\end{thebibliography}\endgroup

\end{document}